\documentclass[showpacs,final,a4paper,pra,10pt,twocolumn]{revtex4}
\usepackage{amsmath}

\input{tcilatex}

\begin{document}

\title{Quantum Key Ditribution Based on Quantum Intensity Correlation of Twin Beams}
\author{Xiaojun Jia, Xiaolong Su, Qing Pan, Kunchi Peng and Changde Xie*}
\affiliation{The State Key Laboratory of Quantum Optics and Quantum
Optics Devices, Institute of Opto-Electronics, Shanxi University,
Taiyuan, 030006, P.R.China}
\date{\today }

\begin{abstract}
A new and simple quantum key distribution scheme based on the
quantum intensity correlation of optical twin beams and the directly
local measurements of intensity noise of single optical beam is
presented and experimentally demonstrated. Using the twin beams with
the quantum intensity correlation of 5dB the effective bit rate of
$2\times 10^7bits/s$ is completed. The noncloning of quantum systems
and the sensitivity of the existing correlations to losses provide
the physical mechamism for the security against eavesdropping. In
the presented scheme the signal modulation and homodyne detection
are not needed.
\end{abstract}

 \maketitle

Quantum key distribution (QKD) usually is a process establishing
secure key between a sender (Alice) and a receiver (Bob) based on
exploiting fundamental properties of quantum mechanics. Then the
random secret key is used for encrypting and decrypting transmitted
secret messages to protect from intercepting of an unauthorized
eavesdropper (Eve). Generally, we say, a communication channel is
secure in the sense that any Eve can be detected by the authorized
communication partners. If Eve is not found the key can be used to
encrypt secret messages, however if Eve is present the key will be
scrapped. According to the fundamental principle of quantum
mechanics any measurement performed on a quantum system will
irreversibly modify it. Therefore, a proper construction of quantum
systems can detect the perturbation introduced by Eve and will allow
secure communication for QKD. Since Bennet and Brassard proposed
first QKD protocol in 1984 a variety of photon-counting QKD systems
have been extensively investigated and experimentally
demonstrated\cite{two,three}. Recently, for developing possibly more
efficient QKD techniques the electromagnetic field amplitudes are
being explored as quantum continuous variables (CV) for QKD by using
non-classical (squeezed or entangled) optical field or
quasi-classical coherent state light\cite
{four,five,six,seven,eight,nine,ten,eleven,twelve,thirteen,fourteen,fifteen,sixteen,seventeen}%
. A QKD protocol based on the transmission of gaussian-modulated
coherent states and shot-noise-limited homodyne detection was
experimentally demonstrated\cite{sixteen}. In almost all proposed
QKD systems of CV the homodyne measurements of quadrature amplitudes
of light field are involved, thus a local oscillation light, which
must be classically coherent with the signal beam, has to be
transmitted along with the signal beam between Alice and Bob. Other
side, the homodyne measurement is relatively complex and more
difficult to be handled than the direct detection of optical beam.
Several attractive entanglement-based QKD schemes of CV relying on
the Einstein-Podosky-Rosen (EPR) quantum correlations of the
quadratures of two-mode squeezed states have been theoretically
proposed\cite {six,seven,ten,twelve}. However, the enhanced security
in these proposals requires high levels of squeezing and low levels
of loss in the channel. Due to some technical difficulties, so far,
there is no any experimental demonstration of QKD based on
entanglement of CV. With respect to the EPR entangled states
resulting from quadrature squeezing\cite {eighteen,nineteen,twenty},
an other bright nonclassical light beams, the intensity correlated
twin beams, can be easily obtained from a nondegenerate optical
parametric oscillator (NOPO) above threshold\cite{twenty one,twenty
two}. Since the injected signal light and thus the phase-locking
between the pump and signal fields are not needed, in addition to
that the exact frequency degenerate between the signal and idler
modes is not required, the experimental sets up generating twin
beams are much simpler and robust than that for EPR entangled
states. The twin beams with the quantum intensity correlation over
$8dB$ have been experimentally produced\cite{twenty one,twenty two},
while the best records of the obtained CV EPR entanglement degree do
not exceed $5dB$ up to now\cite{twenty three,twenty four,Fabre}.
Utilizing the quantum intensity correlation of twin beams, Fabre's
group completed the first experimental demonstration of conditional
preparation of a nonclassical state of light in the CV
regime\cite{twenty five}. The experiment proved that the quantum
intensity correlation between twin beams from NOPO has nonlocal
character. It means, there is the quantum intensity correlation
between output signal and idler light beams of NOPO whatever how far
they are separated. It has been demonstrated, in the ideal case
without losses, the Fourier components of the signal and idler
intensity quantum
fluctuations which lie inside the cavity bandwidth are perfectly correlated%
\cite{twenty six}. Therefore the simultaneously measured
photocurrent fluctuation spectra of signal and idler beams on time
intervals long compared to the cavity storage time should be totally
identical\cite{twenty six}. In a real experiment, the correlation
between the signal and idler photocurrents is not perfect due to the
existence of optical losses. The correlations are characterized
quantitatively by the quantum noise of the intensity difference
between the signal and idler beams which is normalized to the
corresponding shot noise limit (SNL) of the twin beams. In
experiments, the SNL is equal to the quantum noise level of a
coherent state light with same intensity of the twin beams. The
measured intensity difference noise spectrum, normalized to its
associated SNL, is given by\cite {twenty seven}:
\begin{equation}
S(\Omega )=1-\frac{\xi \eta }{1+\Omega ^2}
\end{equation}
where $\Omega $ is the noise frequency normalized to the cavity bandwidth, $%
\eta $ stands for the quantum efficiency of the detection system and
transmission line, and $\xi $ is the OPO output coupling efficiency
depending on the transmission coefficient of the cavity coupling
mirror and the extraneous cavity losses. For given system and
measurement frequency the intensity correlation $S(\Omega )$ has a
certain value, so the quantum correlation is determinant. However,
the instantaneous intensity noise power of each beam (signal or
idler) can randomly change around an average value although the
difference of the noise powers between signal and idler beams almost
keeps unchanging. It means, the noise powers of signal and idler
beams always increase or decrease together whatever how far they
are.
 Using the intensity nonlocal quantum correlation of twin beams
we designed a protocol of CV QKD\ and experimentally demonstrated
its feasibility. By means of Alice's and Bob's local measurements on
own light beam (a half of twin beams) and publicly classical
communication a secure key can be obtained. The quantum no-cloning
principle and the sensitivity of the existing correlations to losses
provide the physical mechanism of the security against eavesdropping
in the presented scheme. The security against the optical tap attack
is analyzed. To the best of our knowledge, this is the first
proposal and experimental demonstration of CV\ QKD based on
utilizing the intensity quantum correlation of twin beams. The
experimental set up is shown in Fig.1. The source of twin beams is a
NOPO consisting of a $\alpha -$cut type-$\Pi $ KTP crystal and a
concave mirror\cite{nineteen,twenty three}. The pump laser of the
NOPO (not shown in Fig.1) is an intracavity frequency-doubled
Nd:YAP/KTP laser. The second harmonic wave output at $0.54\mu m$
wavelength serves as the pump field of the NOPO. The front face of
KTP is coated to be used as the input coupler, the reflectivities of
which are $88\%$ for the pump ($0.54\mu m$) and almost
$100\%$ for the signal and idler beams ($1.08\mu m$). The concave mirror ($%
R=50mm$) used as the output coupler of twin beams is highly reflecting for $%
0.54\mu m$ and its transmission for $1.08\mu m$ is $3\%$. The NOPO
is actively locked on the pump laser by sideband frequency locking
technique through a piezo-electric transducer mounted on the output
coupler. The temperature of the KTP is precisely stabilized within a
mK around the optimal phase-match point to obtain the stable output
intensity of twin beams during whole experimental process. At exact
triple resonance of pump, signal and idler, the oscillation
threshold of NOPO\ is about $3mW$. Under a pump power of $18mW$ the
output twin beams of $6mW$ is obtained and the intensity difference
correlation $S(\Omega )$ measured with a spectrum analyzer in Alice
is $5.0\pm 0.2dB$ below the SNL at $2MHz$ (Fig.2). The corresponding
noise power is $31.6\%$ of SNL and the value of the noise voltage of
the intensity difference ($I_s$) measured with the oscilloscope is
$1.8\pm 0.1mV$ in our system. The signal and idler of the twin beams
with orthogonal polarization are separated by a polarizing beam
splitter (PBS).The signal beam is sent to Bob and the idler beam is
retained by Alice. The photocurrents of two beams are directly
detected by high quantum efficiency photodiodes $D_1$ and $D_2$,
respectively, then is electronically filtered (centre frequency
$2MHz$, $\triangle \Omega =400kHz$) and
amplified. The amplified photocurrents are recorded by the oscilloscopes $%
OS_1$ and $OS_2$ respectively. For keeping the initial correlation
of twin beams at the best extent the parameters of all electronics
devices in signal and idler should be balanced carefully, as well as
the transmission distances and the line losses from NOPO to D1 and
D2 should be adjusted to be same. To establish the right timing of
recording Alice and Bob have to synchronize their clocks and agree
upon a set of time intervals $\triangle t_k$ in which they subdivide
their local measurements. Then, Alice and Bob proceed with a series
of measurements.
At first, Alice and Bob measure the average photocurrents $I_{A0}$ and $%
I_{B0}$ of own beam in a large time interval $\triangle T$ much
longer than the cavity storage time $\triangle t$ ($\triangle
t$=$6.7ns$ for the cavity of our NOPO), respectively. For the case
of balanced intracavity losses of
signal and idler beam, identical transmission lines from NOPO to $D_1$ and $%
D_2$, and same detection electronics, $I_{A0}$ should equal to $I_{B0}$ ($%
I_{A0}$ $=$ $I_{B0}=$ $I_0$ ). However due to the imperfect
intensity correlation, $I_{Ak}$ and $I_{Bk}$ measured at same moment
and within a short time interval $\triangle t_k$ comparable with
$\triangle t$ will have some fluctuation around $I_0$. The
differences of $\left| I_{Ak}-I_{Bk}\right| $ statistically equal to
$I_s$. Therefore, when $I_{Ak}$ is larger (or smaller) than $I_0$
and the value of $\left| I_{Ak}-I_0\right|
$ is larger than $I_s$, $I_{Bk}$ must be also larger (or smaller) than $I_0$%
, vice versa. Before communication, Alice measures $I_0$ and $I_s$
of twin beams generated in her station using two transmission lines
which simulate the real line from Alice to Bob and informs Bob her
measured outcomes publicly. The $I_0$ will become a standard value
and later measured outcomes of $I_{Ak}$ and $I_{Bk}$ will be
normalized to $I_0$. Taking $I_0=0$ for convenience, the values of
$I_{Ak}$ ($I_{Bk}$) larger than $I_0$ are positive and the smaller
values are negative. Then Alice sends the signal beam to Bob and
they perform synchronically local measurements of the photocurrents
at a series of time points $t_k$ ($k=1,2,...$) on own beam. They
record the values $I_{Ak}$ and $I_{Bk}$ as well as the corresponding
measurement time points $t_k$, respectively. For transmitting a
binary key, for example, Alice and Bob agree on that the positive
values of $I_{Ak}$ and $I_{Bk}$ are ''$1$'' and the negative values
are ''$0$''. When $\left| I_{Ak}-I_0\right| \geq I_s$, $I_{Ak}$ and
$I_{Bk}$ must have identical positive or negative symbol. After a
series of local measurement is completed, Alice drops the data of
$\left| I_{Ak}-I_0\right| \leq I_s$ firstly, and publicly tell Bob
''right '' if her measurement outcome at $t_k$ is same with the
secret key she wants to send or ''wrong'' if it is not same. Alice
only announces ''right'' or ''wrong'' but never her measurement
outcomes. For example, if Alice wants to send secret keys of
''$100100010$''
and her measurement outcomes at the time points $t_1$ to $t_{11}$ are ''$%
11(drop)011(drop)0001$'', she publicly informs Bob ''$RWDRRWDRRWW$''
(R--right, W--wrong, D--dropped). If without extraneous disturbance
on signal beam Bob's measurement outcomes should be in agreement
with Alice's that, so he can easily infer the secret keys based on
his measurement outcomes and the public information. Table 1 lists
Alice's and Bob's really experimental measurement outcomes at $t_1$,
$t_2$, ... $t_{14}$. The measured $I_0$ and the $I_s$ are $60.0\pm
0.2mV$ and $1.8\pm 0.1mV$,
respectively. At time points $t_4$, $t_8$, $t_9$, $t_{12}$ and $t_{13}$, $%
\left| I_{Ak}-I_0\right| \leq I_s$ and the corresponding values
should be dropped. Clearly, the symbols of the rest values
respectively measured by Alice and Bob are identical as our
expectation. Based on the outcomes and Alice's public information
Bob can exactly infer any secret keys which Alice wants to send,
such as above-mentioned ''$100100010$''.
Experimental diagram of CV QKD with optical twin beams.
NOPO--Nondegenerate parametric amplifier; F$_{1(2)}$--electronic filter
(centre frequency $\Omega =2MHz$, bandwidth=$400kHz$ ); Am$_{1(2)}$%
--electronic amplifier; PBS--polarizing beam splitter; OS$_{1(2)}$%
--oscillascope; D$_{1(2)}$--photodiode.
In our protocol, one of the
twin beams (idler) is retained in the station of Alice, thus it can
not be disturbed. By comparing the instantaneous noises of two beams
at same moment the presence of Eve can be discovered. Fig.3 shows
the photocurrent fluctuations of $I_{Ak}$ and $I_{Bk}$ respectively
recorded by Alice and Bob at same time interval. The intensity
correlation is obviously presented. Any other presented beam will
not be able to have the same fluctuation at each time point due to
the ''noncloning'' of quantum noise. From Eq.1 we can see, if Eve
uses the optical tap attack the
transmission efficiency of the signal beam must be decreased from initial $%
\eta _0$ to $\eta $ ($\eta <\eta _0$ ). The noise levels of $I_{Bk}$
will increase to $I_{Bk}^{/}$:
\begin{equation}
I_{Bk}^{/}=I_{Bk}+\frac{\xi (\eta _0-\eta )}{1+\Omega ^2}
\end{equation}

For testing the presence of Eve, Alice randomly picks some subsets
of the measured data to be the test sets and the rest to be the key.
After a transmission is completed Alice announces the noise levels
and the noise patterns of the test subsets on a public channel. Bob
compares the values and patterns with that measured by himself at
same time points one by one. If all are in reasonable agreement with
that measured by Alice, Bob can be sure that the transmission is
secure and the rest unclosed data can be used for the secret key. If
most of $\left| I_{Ak}-I_{Bk}\right| $ are larger than $I_s$ and the
patterns of intensity fluctuation are not correlated, they have to
discard the transmission and try again since an eavesdropper may
have been present.
 For conclusion, a CV QKD scheme using quantum
intensity correlation is presented. In this scheme, the signal
modulation of quadratures of optical field and the homodyne
detection are not needed, so the system is significantly simplified
with respect to most proposed CV\ QKD protocols\cite
{four,five,six,seven,eight,nine,ten,eleven,twelve,thirteen,fourteen,fifteen,sixteen}%
. The bit transmission rate is only limited by the cavity storage
time and the response time of electronics which usually is the order
of $10^7bits/s$, thus the high bit rate is available for the
presented protocol. In this experiment, we recorded $5\times 10^4$
data within $1ms$, in which about
40\% of the measured $\left| I_{Ak}-I_0\right| $ values are smaller than $%
I_s $ and have to be dropped, so the effective bit rate is about
$2\times 10^7bits/s$. An other advantage of the presented scheme is
that the limited and experimentally reachable quantum correlation of
twin beams may be used for establishing secure key.
Acknowledgements: This work was supported by the National Natural
Science Foundation of China(Grant No.60238010, 60378014).

Captions of figures:

Fig.1 Experimental diagram of CV QKD with optical twin beams.
NOPO--Nondegenerate parametric amplifier; F$_{1(2)}$--electronic
filter
(centre frequency $\Omega =2MHz$, bandwidth=$400kHz$ ); Am$_{1(2)}$%
--electronic amplifier; PBS--polarizing beam splitter; OS$_{1(2)}$%
--oscillascope; D$_{1(2)}$--photodiode.

Fig.2 The intensity difference correlation noise power measured by
Alice at 2MHz as a function of time. 1- SNL; 2-The intensity
difference correlation
noise power. The measurement parameters of SA: RBW(Resolution Band Width)-$%
100kHz$; VBW(Video Band Width)-$100Hz$.

Fig.3 Photocurrent fluctuation patterns of $I_{Ak}$ and $I_{Bk}$
respectively recorded by OS$_1$ and OS$_2$ during $5\mu s$.

Table.1 $I_{Ak}$ and $I_{Bk}$ normalized to $I_0$. ($I_0$= $60.0\pm 0.2mV$, $%
I_s$= $1.8\pm 0.1mV$, $\triangle t_k$= $20ns$)


\begin{thebibliography}{99}
\bibitem{two}  C. H. Bennett, G. Brassard, Proceeding of IEEE Inernational
Conference on Computers, System and Signal Processing, Bangalore,
India, P.175 (1984)

\bibitem{three}  N. Gissin et al, Rev. Mod. Pys. 74, 145 (2002)

\bibitem{four}  S. L. Braunstein, A. K. Pati, Quantum Information with
Continuous Variables, (Kluwer Academic Publishers, 2003)

\bibitem{five}  M. Hillery, Phys. Rev. A61, 022309 (2000)

\bibitem{six}  T. C. Ralph, Phys. Rev. A61, 010303(R) (2000); ibid 62,
062306 (2000)

\bibitem{seven}  M. D. Reid, Phys. Rev. A62, 062308 (2000)

\bibitem{eight}  D. Gottesman J. Preskill, Phys.ev. A63, 022309 (2001)

\bibitem{nine}  N. Cerf, M. Levy, G. Van Assche, Phys. Rev. A 63, 052311
(2001)

\bibitem{ten}  K. Bencheikh et al, J. Mod. Opt. 48, 1903 (2001)

\bibitem{eleven}  N. Cerf, S. Iblisdir, G. Van Assche, Eur. Phys. J. D18,
211 (2002)

\bibitem{twelve}  Ch. Silberhorn, N. Korolkova, G. Leuchs, Phys. Rev. Lett.
88, 167902 (2002)

\bibitem{thirteen}  M. Osaki, M. Ban, Phys. Rev. A68, 022325 (2003)

\bibitem{fourteen}  Ch. Silberhorn et al, Phys. Rev. Lett. 89, 167901 (2002)

\bibitem{fifteen}  F. Grosshans, Ph. Grangier, Phys. Rev. Lett. 88, 057902
(2002)

\bibitem{sixteen}  F. Grosshans et al, Nature(London) 421, 238 (2003)

\bibitem{seventeen}  Ch. Weedbrook et al, Phys. Rev. Lett. 93, 170504 (2004)

\bibitem{eighteen}  A. Furusawa et al, Science 282,706 (1998)

\bibitem{nineteen}  X. Y. Li et al, Phys. Rev. Lett. 88, 047904 (2002)

\bibitem{twenty}  H. Yonezawa, T. Aoki, A. Furusawa, Nature(London) 431, 430
(2004)

\bibitem{twenty one}  J. Mertz et al, Opt. Lett. 16, 1234 (1991)

\bibitem{twenty two}  J. R. Gao et al, Opt. Lett. 23, 870 (1998)

\bibitem{twenty three}  X. J. Jia et al, Phys. Rev. Lett. 93, 250503 (2004)

\bibitem{twenty four}  N. Takei, H. Yonezawa, T. Aoki, A. Furusawa, arXiv:
quant-ph/0501086 (2005)

\bibitem{Fabre}  J. Laurat et al, Phys. Rev. A 71, 022313 (2005)

\bibitem{twenty five}  J. Laurat et al, Phys. Rev. Lett. 91, 213601 (2003)

\bibitem{twenty six}  S. Reynaud, C. Fabre, E. Giacobino, J. Opt. Soc. Am.
B4, 1520 (1987)

\bibitem{twenty seven}  C. Fabre et al, J. de Physique 50, 1209 (1989)
\end{thebibliography}
\end{document}